\documentclass{aastex631}
\usepackage{psfig,epsfig,amsfonts,amsmath,amssymb,graphicx,morefloats,appendix,tensor}
\usepackage{color}
\usepackage{natbib}
\usepackage{hyperref}
\usepackage{subfigure}
\renewcommand{\dataset}[1]{doi:{#1}, \url{https://doi.org/#1}}

\newcommand{\appropto}{\mathrel{\vcenter{
  \offinterlineskip\halign{\hfil$##$\cr
    \propto\cr\noalign{\kern2pt}\sim\cr\noalign{\kern-2pt}}}}}

\begin{document}

\title{Separating flare and secondary atmospheric signals with RADYN modeling of near-infrared JWST transmission spectroscopy observations of TRAPPIST-1}

\author[0000-0002-0583-0949]{Ward S. Howard}
\affil{Department of Astrophysical and Planetary Sciences, University of Colorado, 2000 Colorado Avenue, Boulder, CO 80309, USA}
\affil{NASA Hubble Fellowship Program Sagan Fellow}

\author[0000-0001-7458-1176]{Adam F. Kowalski}
\affil{Department of Astrophysical and Planetary Sciences, University of Colorado, 2000 Colorado Avenue, Boulder, CO 80309, USA}
\affil{National Solar Observatory, University of Colorado Boulder, 3665 Discovery Drive, Boulder, CO 80303, USA}
\affil{Laboratory for Atmospheric and Space Physics, University of Colorado Boulder, 3665 Discovery Drive, Boulder, CO 80303, USA.}

\author[0000-0002-3328-1203]{Michael Radica}
\affil{Department of Astronomy \& Astrophysics, University of Chicago, 5640 South Ellis Avenue, Chicago, IL 60637, USA}
\affil{NSERC Postdoctoral Fellow}

\author[0000-0001-6362-0571]{Laura Flagg}
\affil{Department of Physics and Astronomy, Johns Hopkins University, 3400 N Charles St, Baltimore, MD 21218, USA}

\author[0009-0009-3020-3435]{Valeriy Vasilyev}
\affil{Max Planck Institute for Solar System Research, Justus-von-Liebig-Weg 3, 37077 G{\"o}ttingen, Germany}

\author[0000-0002-3627-1676]{Benjamin V. Rackham} 
\affil{Department of Earth, Atmospheric and Planetary Sciences, Massachusetts Institute of Technology, Cambridge, MA 02139, USA}
\affil{Kavli Institute for Astrophysics and Space Research, Massachusetts Institute of Technology, Cambridge, MA 02139, USA}

\author[0000-0001-5455-6678]{Guadalupe Tovar Mendoza}
\affil{Department of Physics and Astronomy, Johns Hopkins University, 3400 N Charles St, Baltimore, MD 21218, USA}
\affil{NSF MPS-Ascend Postdoctoral Fellow}

\author[0000-0001-7891-8143]{Meredith A. MacGregor}
\affil{Department of Physics and Astronomy, Johns Hopkins University, 3400 N Charles St, Baltimore, MD 21218, USA}

\author[0000-0002-8842-5403]{Alexander I. Shapiro}
\affil{Max Planck Institute for Solar System Research, Justus-von-Liebig-Weg 3, 37077 G{\"o}ttingen, Germany}

\author[0000-0003-4844-9838]{Jake Taylor}
\affil{Department of Physics, University of Oxford, Parks Rd., Oxford OX1 3PU, UK}

\author[0000-0002- 2195-735X]{Louis-Philippe Coulombe}
\affil{Institut Trottier de recherche sur les exoplan{\`e}tes and D{\'e}partement de Physique, Universit{\'e} de Montr{\'e}al, 1375 Avenue Th{\'e}r{\`e}se-Lavoie-Roux, Montr{\'e}al, QC, H2V 0B3, Canada}

\author[0000-0003-4676-0622]{Olivia Lim}
\affil{Institut Trottier de recherche sur les exoplan{\`e}tes and D{\'e}partement de Physique, Universit{\'e} de Montr{\'e}al, 1375 Avenue Th{\'e}r{\`e}se-Lavoie-Roux, Montr{\'e}al, QC, H2V 0B3, Canada}

\author[0000-0002-6780-4252]{David Lafreni{\`e}re}
\affil{Institut Trottier de recherche sur les exoplan{\`e}tes and D{\'e}partement de Physique, Universit{\'e} de Montr{\'e}al, 1375 Avenue Th{\'e}r{\`e}se-Lavoie-Roux, Montr{\'e}al, QC, H2V 0B3, Canada}

\begin{abstract}
Although TRAPPIST-1's temperate planets have the highest transmission signals of any known system, flares contaminate 50--70\% of transits at the 1000~ppm level, far above 100 ppm secondary atmospheric signals. Efforts to mitigate flare contamination and assess impacts on radiation environments are each hampered by a lack of empirical spectral analysis and physics-based modeling. We present spectrotemporal analysis and radiative-hydrodynamic modeling of 5.5 hr of NIRISS and NIRSpec observations of six TRAPPIST-1 flares of 2.2--8.7$\times$10$^{30}$\,erg. Flare lines and continua are characterized using grid searches of \texttt{RADYN} beam-heating models spanning 10$^4\times$ in electron beam parameters. Best-fit models indicate these flares result from moderate-intensity beams with emergent electron fluxes of $F_e$=10$^{12}$\,erg\,s$^{-1}$~cm$^{-2}$ and energies $\leq$37~keV, although all models over-predict the Paschen jump. These models predict XUV, FUV, and NUV counterparts to the infrared peak fluxes of 8.9--28.9$\times$10$^{27}$, 4.3--13.9$\times$10$^{26}$, and 3.4--11.4$\times$10$^{27}$\,erg\,s$^{-1}$, respectively. Scaling the flare rate into the XUV suggests flaring contributes 1.35$_{-0.15}^{+2.0}\times$ quiescence~yr$^{-1}$. We bin integrations of similar flare effective temperature to construct fiducial flare spectra from 2000--4500~K in order to develop separate empirical and \texttt{RADYN}-based mitigation pipelines. Both pipelines are applied to all 5.5 hr of $R$=10 data, resulting in maximum residuals from 1--2.8\,$\mu$m of 100--140~ppm and typical residuals of 54$\pm$14 and 65$\pm$17 ppm for the empirical and \texttt{RADYN}-based pipelines, respectively. Injection testing supports 3$\sigma$ detection capability for CO$_2$ atmospheres with features of 150--250 ppm, with weak evidence (BF$\approx$3) still obtained at 130~ppm. Our results motivate multi-wavelength observations to improve model fidelity and test high-energy predictions.
\end{abstract}


\section{Introduction}\label{sec:intro}
Stellar activity is the primary bottleneck for detecting and interpreting secondary atmospheres in transmission on terrestrial exoplanets \citep{deWit:2024}. All terrestrial planets with a high transmission spectroscopy metric (TSM) orbit M dwarfs and are subject to extreme and highly-variable radiation environments and strong transit light source (TLS) effects \citep{Kempton:2018, Rackham:2023, Paudel:2024}. Perhaps no system illustrates the challenge and opportunity for stellar characterization and mitigation more so than \object{TRAPPIST-1}, a nearby M7.5 ultracool dwarf (UCD) with three temperate (i.e. in the so-called ``habitable zone") terrestrial planets and seven terrestrial planets total \citep{Gillon:2017}. JWST has invested $>$260 hr into transit spectroscopy observations of the system, widely considered our best chance to characterize temperate secondary atmospheres due to the lack of comparable systems of TSM$\gtrsim$20 \citep{deWit:2024}. The high-energy emission \citep{Wheatley:2017, Becker:2020} drives variable atmospheric escape and disequilibrium chemistry \citep{Hu:2020, Ranjan:2023}, calling into question the applicability of time-independent models often used to plan and interpret transmission spectroscopy observations (e.g. \citealt{Lustig-Yaeger:2019, Lincowski:2023, Krissansen-Totton:2024}). At longer wavelengths of 1--5\,$\mu$m, slowly-varying contamination from starspots and faculae ($t_\mathrm{spot}$ of hr--d) produce signals of 200--700 ppm \citep{Lim:2023, deWit:2024, Berardo:2024}. Rapid increases up to 2000 ppm relative to this baseline occur during stellar flares at 1--5\,$\mu$m \citep{Howard:2023}, exceeding the 50—100 ppm signals of 0.1--10 bar CO$_2$ terrestrial atmospheres by more than an order of magnitude \citep{Lustig-Yaeger:2019}. While significant attention has been devoted to mitigation of slowly-varying contamination (e.g., \citealt{Lim:2023, Berardo:2024, Radica:2024b, Rathcke:2025}), flare mitigation remains in infancy due to a lack of empirical characterization and modeling \citep{deWit:2024}. 

The stochastic nature of flares combined with their rapid evolution in both time and wavelength dimensions provide unique challenges for mitigation. Flare contamination has now been observed during JWST transit spectroscopy for a range of K7--M7 dwarfs, with the majority from TRAPPIST-1 (e.g. \citealt{Howard:2023, Piaulet:2024, Radica:2024b, Murray:2025, Espinoza:2025}). Initial mitigation approaches relied upon scaling contaminated spectra by the H$\alpha$ time series \citep{Lim:2023, Radica:2024b}. However, the two-dimensional spectrotemporal flare evolution is not captured by this approach, resulting in high residuals \citep{Radica:2024b}. Fitting and subtracting a one-component Planck function to each flare-affected spectrum is the simplest 2D approach. \citet{Howard:2023} used Planck fitting to detrend a transit of TRAPPIST-1f during a large flare, reporting residuals of $\sim$400 ppm. 

Accurate physics-based flare models are likely needed for the robust detection of secondary atmospheres \citep{Howard:2023}, especially when confounding signals from spots, faculae, and flares are all present. Recent progress in the development of time-dependent radiative-hydrodynamic (RHD) flare models has enabled the NUV/optical emission of a wide range of solar and stellar flares to be successfully reproduced \citep{Kowalski:2024}. However, model fidelity has not yet been demonstrated for the infrared spectra of stellar flares, nor for their connection to X-ray--UV counterparts. These models must explain the trends that are beginning to emerge from the JWST flares, including the prevalence of low effective peak temperature ($T_\mathrm{eff}<$5000 K), a fast-rise, exponential decay in flare $T_\mathrm{eff}$ that evolves on sub-minute timescales and lasts 0.5--2 hr, and spectra dominated by continuum rather than line emission \citep{Howard:2023, Piaulet:2024}.

We capitalize upon the growing sample of TRAPPIST-1 flares to identify commonalities in flare spectra of a given $T_\mathrm{eff}$, and to investigate model fidelity for a broad grid of time-dependent flare heating models generated with the \texttt{RADYN} code \citep{Carlsson_Stein:1995, Carlsson_Stein:1997} and presented in \citet{Kowalski:2024}. In particular, the increase in the NIRISS flare sample from GO 2589 (PI: O. Lim) and GTO 1201 (PI: D. Lafreni{\`e}re) since \citet{Howard:2023} now enables separate estimates of flare heating environments from the continuum and lines across multiple events. In this work, we use the updated sample of six NIRISS/NIRSpec flares to construct fiducial-$T_\mathrm{eff}$ flare spectra and perform \texttt{RADYN} grid searches against the continuum and line emission. Using the best-fit models, we investigate the high-energy counterparts and suitability of \texttt{RADYN} for flare mitigation. In \S \ref{jwst_flare_obs}, we describe the observations and data reduction. In \S \ref{fiduc_teff}, we describe the fiducial spectra. In \S \ref{empirical_approach}, we present an empirical mitigation pipeline based on fiducial spectral features. In \S \ref{rhd_modeling}, we present the \texttt{RADYN} grid search and model-based mitigation pipeline. In \S \ref{escape_photochem}, we estimate the contribution of X-ray--UV counterparts to the radiation environment. In \S \ref{inj_recov}, we perform injection-and-recovery tests of secondary atmospheres. Finally, in \S \ref{discuss_conclude}, we discuss the implications of the empirical and model analyses for stellar radiation environments and mitigation, and conclude.

\begin{figure*}
	\centering
	{
		\includegraphics[width=0.99\textwidth]{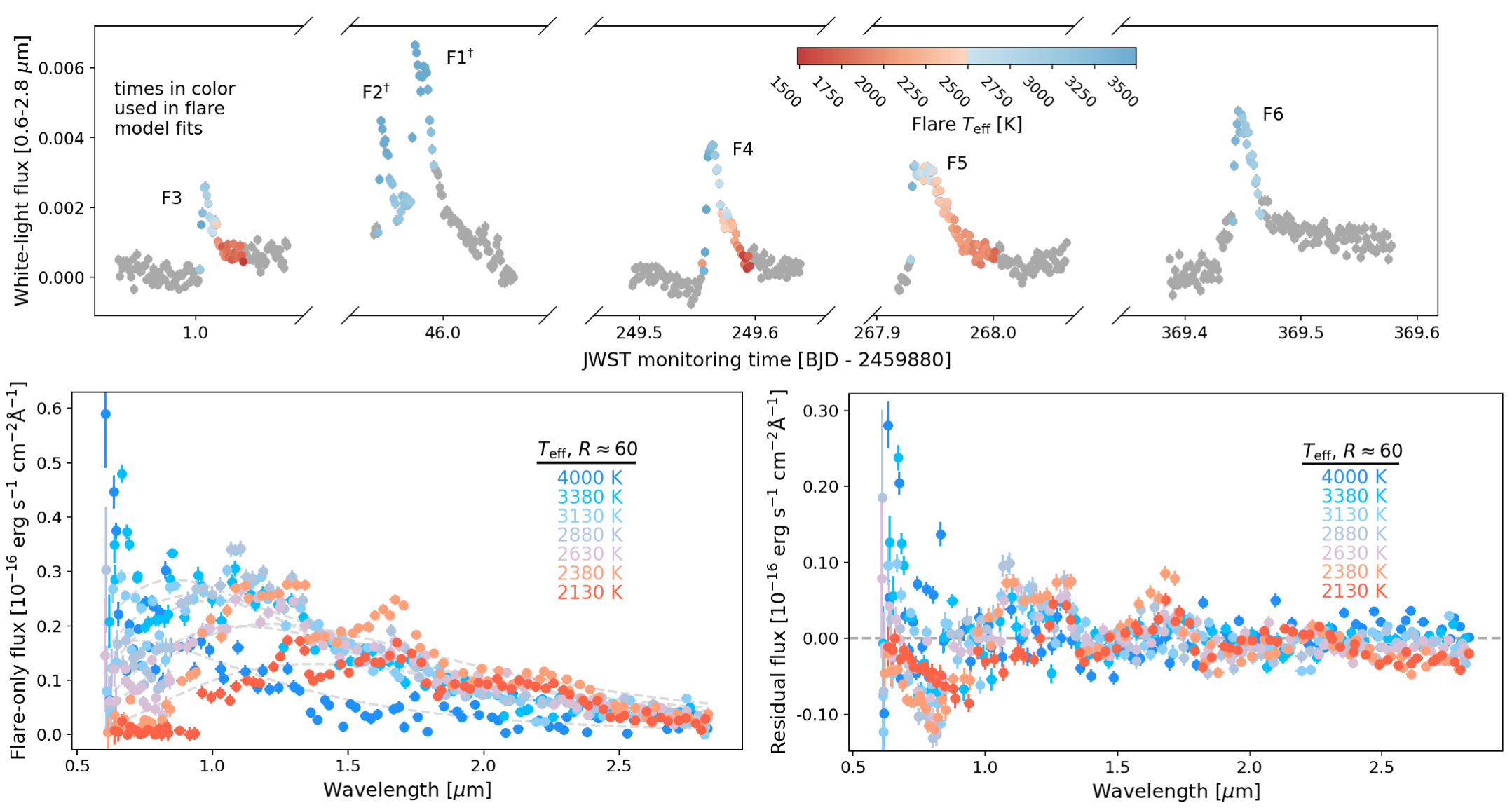}
	}
	\caption{Top: Wavelength-integrated (0.6--2.83\,$\mu$m), transit-normalized flare light curves of our sample. Flare temperatures of integrations used for model validation and mitigation are scaled by color. Bottom left: Fiducial-$T_\mathrm{eff}$ flare spectra are computed from similar-temperature integrations. Best-fitting blackbody curves are shown with dashed gray lines. Bottom right: Quiescent features appear in the residual spectra after subtracting the best-fit blackbody curves.}
	\label{fig:activity_overview}
\end{figure*}

\section{JWST observations of the TRAPPIST-1 flare sample}\label{jwst_flare_obs}
We include two observations of TRAPPIST-1c obtained with NIRISS SOSS as part of GO 2589 (PI: O. Lim) and five observations of TRAPPIST-1f with NIRISS as part of GTO 1201 (PI: D. Lafreni{\`e}re). A NIRSpec BOTS PRISM observation of TRAPPIST-1g from GO 2589 is also included due to the presence of two flares previously analyzed in \citet{Howard:2023}. Our NIRISS sample contains a total of 923 integrations obtained at 104\,s cadence, for a combined 26.66\,hr of flare monitoring. We include 6,570 NIRSpec integrations obtained at 1.6\,s cadence during the flares, which we bin to match the NIRISS 104\,s cadence, for a combined 99 binned integrations or 2.86\,hr. Although the 2022-10-28 visits were previously analyzed in \citet{Howard:2023}, we re-reduce and analyze them to ensure a uniform analysis across the sample. A summary of the observations is given in \autoref{table:observations}.

\begin{table}
\centering
\caption{Overview of TRAPPIST-1 Observations used in this Work}
\begin{tabular}{ccccccccccc}
\hline
\hline
Program & Obs. & Planet & Visit & Date & $t_\mathrm{start}$ & Date & $t_\mathrm{end}$ & $\Delta t_\mathrm{target}$ & $N_\mathrm{integ}$ & Flares? \\
        &      &        &       & ISO  & UTC                & ISO & UTC              & hr                         &         &  \\
\hline
2589 & 1 & b & 1 & 2022-07-18 & 14:00:06 & 2022-07-18 & 19:04:38 & 4.41 & 153 & no \\
2589 & 2 & b & 2 & 2022-07-20 & 02:15:42 & 2022-07-20 & 07:20:14 & 4.41 & 153 & yes (F7) \\
2589 & 3 & c & 1 & 2022-10-28 & 18:14:52 & 2022-10-28 & 23:29:50 & 4.58 & 159 & no \\
2589 & 4 & c & 1 & 2023-10-31 & 20:16:59 & 2023-11-01 & 02:06:43 & 4.59 & 159 & F6 \\
1201 & 101 & f & 1 & 2022-10-28 & 10:14:49 & 2022-10-28 & 14:04:50 & 3.50 & 121 & F3 \\
1201 & 102 & f & 1 & 2023-07-03 & 23:08:18 & 2023-07-04 & 03:16:46 & 3.50 & 121 & no \\
1201 & 103 & f & 1 & 2023-07-22 & 09:05:11 & 2023-07-22 & 13:25:36 & 3.50 & 121 & no \\
1201 & 104 & f & 1 & 2023-06-24 & 18:37:49 & 2023-06-24 & 22:54:08 & 3.50 & 121 & F5 \\
1201 & 105 & f & 1 & 2023-06-15 & 12:35:02 & 2023-06-15 & 17:24:06 & 3.50 & 121 & F4 \\
2589 & 6 & g$^\dag$ & 1 & 2022-07-17 & 04:41:12 & 2022-07-17 & 09:58:17 & 4.95 & 11117 & no \\
2589 & 7 & g$^\dag$ & 1 & 2022-12-12 & 09:55:32 & 2022-12-12 & 15:09:00 & 4.95 & 11117 & (F1\&2) \\
\hline
\hline
\end{tabular}
\label{table:observations}
{\newline\newline \textbf{Notes.} Description of the observations used in this work. Visits without $\geq$10$^{30}$ erg flares are only used for flare rate measurements. Columns are the program ID, JWST Visit Status Report (VSR) observation, planet, visit sequence number, date, start and end times listed for archived observations in the VSR, on-target science time, number of integrations, and the presence of flares. A dagger denotes the instrument and mode as NIRSpec BOTS PRISM, while none denotes NIRISS SOSS.}
\vspace{0.05cm}
\end{table}

\subsection{Reduction of JWST observations and extraction of flare spectra and light curves}\label{reduct_pipe}
NIRISS SOSS time series observations are reduced using the \texttt{exoTEDRF} package \citep{Feinstein:2023, Radica:2023, Radica:2024a}. Briefly summarizing, \texttt{exoTEDRF} performs differential calibration of the raw data, applying bias and 1/f noise correction at the group level within each integration. Flat-field and bad pixel correction are applied and the background is subtracted from each image using the SUBSTRIP256 SOSS zodiacal model provided by STScI \citep{Rigby:2023}, which is scaled separately to the median image flux of two comparison regions above and below the pick-off mirror jump \citep{Lim:2023, Radica:2024c}. Detector-level effects such as trace drift or tilt events are identified using principle component analysis applied to the 2D detector images as described in \citet{Radica:2024b}. Finally, one-dimensional spectra are extracted using a simple box aperture of 40 pixels. 

The spectrum of each integration is flux-calibrated using a 1D calibration vector constructed by comparing a SOSS spectrum of an A0 star to either a model spectrum of that stellar type or a reliable observed spectrum in physical units. The calibration vector is adjusted for the position of the pupil wheel at the time of the TRAPPIST-1 observation versus the A0 star. The wavelength solution of NIRISS can vary by several pixels due to the vertical position of the trace. Wavelength corrections are obtained by comparison of the expected positions and intensities of the strong hydrogen lines identified in \citet{Howard:2023} against the observed positions in the flare peak integrations. NIRSpec data are reproduced from \citet{Howard:2023}.

Extracted spectra are convolved with gray (0.6--2.83\,$\mu$m) and TESS (0.6--1.0\,$\mu$m) response functions and wavelength-integrated to create white-light and TESS-band  light curves, respectively. Spectra are transit normalized, where best-fit transit parameters are identified using a joint transit and flare model fit to the white-light light curves as described in the Appendix. Flare light curves are obtained by subtracting the best-fit transit models from the white-light and TESS-band light curves in order to identify in-flare integrations and measure flare energies following \citet{Howard:2023}. We limit the present analysis to six flares of $E_\mathrm{TESS}\geq$10$^{30}$\,erg (SNR$\geq$11.4) as temperatures of smaller flares are not well-constrained at 104\,s cadence. The normalized flux amplitudes of the NIRISS flares range from 0.28--0.48\%, while the amplitudes of the NIRSpec flares are 0.45 and 1.31\%. These values are broadly consistent with upper-limits on 2MASS flare amplitudes \citep{Tofflemire:2012}. The flare-only spectrum (hereafter, flare spectrum) of each integration is measured by subtracting the average spectrum of the out-of-flare integrations on a per-visit basis to minimize differences in surface inhomogeneities. Furthermore, out-of-flare integrations in each visit are selected to avoid times in obvious trends in the white-light light curves. However, we assess the impact of quiescent variability by splitting the out-of-flare integrations in half and comparing the resulting $T_\mathrm{eff}$ time series of the NIRISS flare-only spectra when using a preflare spectrum constructed from the earlier versus later preflare times. The two $T_\mathrm{eff}$ time series differ by 6--7\% when using these two preflare reference spectra, and the temperatures from each reduction occur within the errorbars of the other one. Effective temperature time series are computed from the flare spectrum of each integration using a single-component Planck model of temperature $T_\mathrm{eff}$ and filling factor $X_\mathrm{eff}$ following \S 5.2 of \citet{Howard:2023} and are shown in the Appendix.

\section{Construction of fiducial temperature flare spectra}\label{fiduc_teff}
The white-light flux and flare temperature time series are used to select a sample of integrations during good time intervals (GTIs) for the construction of fiducial flare spectra and \texttt{RADYN} modeling. We define GTIs as a continuous series of integrations for which the FWHM-duration flux of the white-light light curve exceeds 5$\sigma$ above the local noise and all $T_\mathrm{eff}$ measurements exceed SNR=3, although we note 99.5\% of these temperature measurements exceed SNR=5. An additional requirement is imposed that both the white-light and temperature time series follow a fast-rise, exponential decay profile consistent with the impulsive phase of a flare \citep{Kowalski:2013}. The slow decay phase with its low signal and complicated $T_\mathrm{eff}$ evolution is left for future work. Our final sample is composed of 189 flare-contaminated integrations during the GTIs, equivalent to 5.5 hr or $N_\mathrm{transit}\approx6$ transit durations. White-light flare light curves are shown in the top panel of \autoref{fig:activity_overview}, where $T_\mathrm{eff}$ is shown in color and non-GTI times in gray.

The $T_\mathrm{eff}$ time series is then used to identify integrations from separate flares with comparable $T_\mathrm{eff}$ values, enabling binning of integrations with similar $T_\mathrm{eff}$ from different flares to construct fiducial temperature spectra. Bins of 250\,K are selected from 2000--3500\,K to ensure $\sim$20 points per bin so that binned spectra can be compared at the 4.5$\sigma$ level. The uncertainty on the mean spectral flux difference per resolution element between template spectra for the different $T_\mathrm{eff}$ bins is always sufficient for a 3$\sigma$ measurement, but not always sufficient for a 5$\sigma$ one. We therefore report the lowest uncertainty across all data, e.g. 4.5$\sigma$. All integrations of 3500--4500\,K are binned together to ensure 20 points per bin given the rarity of these points. The 1$\sigma$ variation in the integrations' spectra about the mean spectrum is 1.7--4 ppt. We also verify the 1$\sigma$ range of the filling factors within each bin is $\leq$0.1\%, which enables stacking of the individual spectra in the bins. The best-fit temperatures to these template spectra are 4010, 3385, 3135, 2885, 2620, 2385 and 2115 K, and the associated filling factors are 0.03, 0.11, 0.14, 0.21, 0.27, 0.44, and 0.51\%, respectively.

The lower panels of \autoref{fig:activity_overview} show the fiducial spectrum for each temperature bin and the residual structure after subtraction of a blackbody of $T_\mathrm{eff,~bin}$, whose fits are shown as dashed gray lines. The increased SNR of the binned spectra reveals undulating structure with larger amplitudes for lower $T_\mathrm{eff,~bin}$. We attribute this structure to imprinting of the quiescent stellar spectrum, as the positions of the residual peaks align with the dominant molecular absorption features (e.g. H$_2$O) of the stellar SED at these wavelengths \citep{Wilson:2021}. Differences in the brightness and optical thickness of the lower stellar atmosphere between preflare and flare states can imprint on the flare-only spectrum, especially given high quiescent brightness of TRAPPIST-1 at infrared wavelengths. This results from the implicit assumption in the pre-flare spectral subtraction that changes are due to the flare \citep{Kowalski:2013}, similar to the assumption of spectral similarity between the stellar disk and transit chord in the TLS effect \citep{Rackham:2018}. \citet{Vasilyev:2025} argue the quiescent structure results from the disappearance of a surface magnetic feature resulting from the flare and therefore is not part of the flare spectrum per se. Unrecognized instrumental time-dependent systematic trends such as those reported in NIRSpec PRISM \citep{Rustamkulov:2023} and NIRISS SOSS \citep{Feinstein:2023} data of WASP-39b suggest another possible origin. However, we remain agnostic on the feature origins and include all spectral changes during the flares in our flare-only spectra.

\begin{figure*}
	\centering
	{
		\includegraphics[width=0.99\textwidth]{new_jwst_peakflarespec_radyn.png}
	}
	\caption{Top: Peak flare spectrum of each event overlaid with the best-fit \texttt{RADYN} model, where fits are color-coded by reduced $\chi^2$ value. The full model grid is shown for reference in gray, where each gray model is scaled by the filling factor of the highlighted model. Models qualitatively describe line and 0.82--2.83\,$\mu$m continuum emission, but predict a large Paschen jump not present in the observations. Daggers denote NIRSpec flares. Bottom: A close-up of the best-fit \texttt{RADYN} line models for the F4 event compared with the NIRISS data. Although not included in \texttt{RADYN}, P$\gamma$ is shown to alleviate blending concerns for the He$_\mathrm{I}$ IRT.}
	\label{fig:peak_spectra}
\end{figure*}

\begin{figure*}
	\centering
	{
		\includegraphics[width=0.99\textwidth]{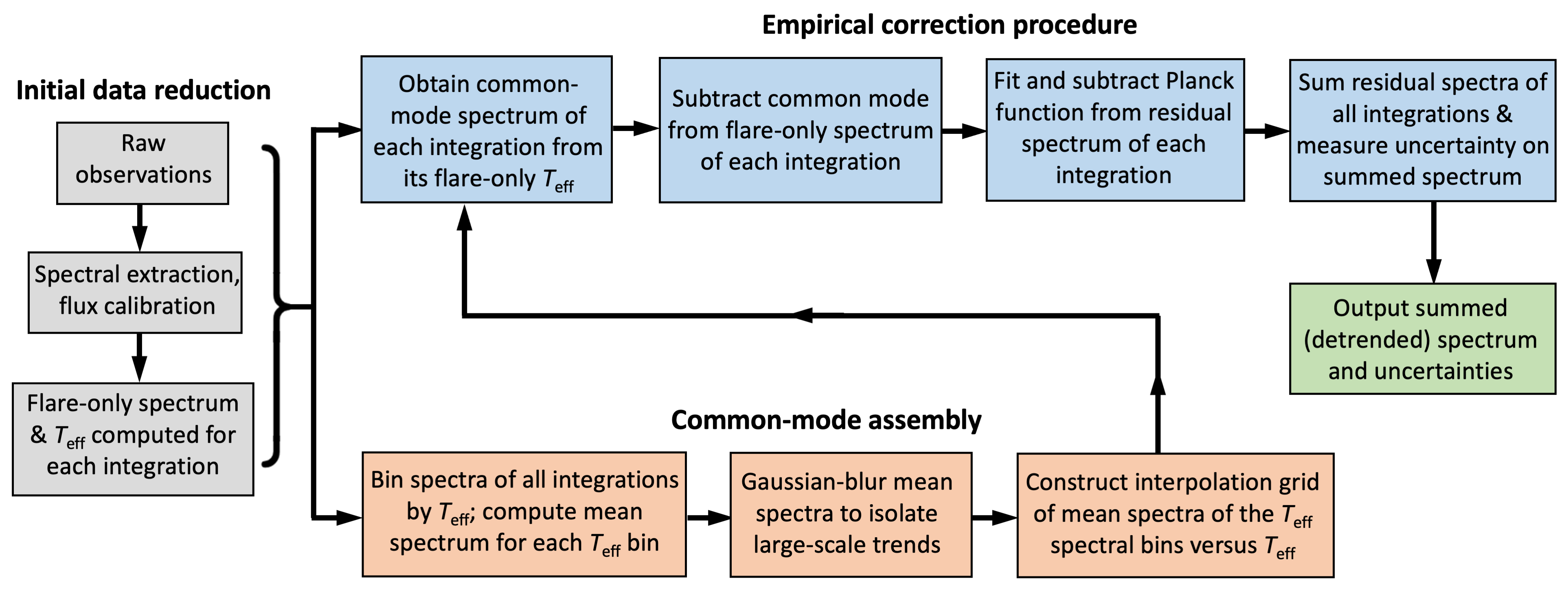}
	}
	\caption{Flowchart illustrating the empirical flare correction procedure. We note the steps in the physics-based procedure are similar to those in the empirical procedure.}
	\label{fig:empirical_pipedemo}
\end{figure*}

\section{Empirical flare modeling with fiducial temperature templates}\label{empirical_approach}
We develop an empirical mitigation pipeline based on the observation that fiducial flare spectral residuals in \autoref{fig:activity_overview} share similar features that evolve with temperature. The similarity in these low-level spectral signals motivates the adoption of the fiducial spectral residuals as common-mode spectra, or shared signals within each temperature bin. We outline our empirical mitigation procedure in \autoref{fig:empirical_pipedemo}, which begins with inputs of the flare-only spectrum and initial $T_\mathrm{eff}$ measurement of each integration and proceeds along two parallel tracks in order to create a summed output spectrum consisting of the detrended spectra of all integrations. The first track is assembly of the common-mode interpolation grid (shown in \autoref{fig:common_modes}).

In the first track, common-mode spectra are computed for the mean spectrum in each fiducial temperature bin by subtracting the best-fit blackbody function, where the temperature is fixed by the bin and the filling factor is a free parameter. Filling factors range from 0.02--0.6\% for integrations above 2400 K, increasing as an inverse function of temperature as observed in \citet{Howard:2023}. The residual spectrum is blurred using a Gaussian kernel of 0.006\,$\mu$m for the second NIRISS order and 0.04\,$\mu$m for the NIRISS first order (and for NIRSpec) to isolate smoothly-varying residual trends. Uncertainties of 0.5--2 ppt on the $T_\mathrm{eff}\leq$3500\,K 1--2.8\,$\mu$m common-mode spectra are propagated from the individual flux errors. Any correlations induced by the blurring process are not explicitly propagated into the uncertainty estimates, but are ultimately accounted for in the residuals reported for the common-mode subtracted spectra. Few $T_\mathrm{eff}>$3500\,K values exist, so uncertainties are 1--4 ppt. The common-mode spectra of the $T_\mathrm{eff}$ bins are assembled into an interpolation grid and $T_\mathrm{eff}$ values measured for the flare spectra of all integrations, so that an interpolated common-mode spectrum can be subtracted from that integration's spectrum. 

The final step in the empirical correction pipeline is subtraction of a new Planck fit to the common-mode detrended spectrum of each integration to produce a residual spectrum. Common-mode spectra are computed separately for NIRISS and NIRSpec integrations since several $\sim$5 ppt level deviations between the NIRISS and NIRSpec residual spectra exist, suggesting a mix of astrophysical and instrumental origins. The clearest of these deviations is a consistent dip at 1.4\,$\mu$m in the NIRSpec common modes. Each common mode for the bins of $T_\mathrm{eff}\leq$3500\,K is computed from the median spectrum of $\sim$10--40 integrations, minimizing the contribution of any one integration. This ensures the improvement in flare mitigation following subtraction of the common mode from the original spectrum is not due to the structure of that particular spectrum. However, the common mode of the $T_\mathrm{eff}\leq$3500--4500 K NIRISS bin is computed from the median of only four integrations given the rarity of these temperatures in the NIRISS flare data. While this is likely to artificially improve mitigation for these four integrations, it does not impact the overall mitigation computed over 189 integrations and will decrease as the number of flares available for empirical correction continues to grow. We also swap adjacent common-mode spectral bins to test sensitivity to flare $T_\mathrm{eff}$ at the $\sim$250 K level. While the 1.2--2.5$\mu$m residuals using the swapped modes remain unchanged, they are 2--4$\times$ larger at the edges of the spectral range. Further details on the common-mode spectra are given in the Appendix.

In the second track, a common-mode spectrum is constructed for each integration from its $T_\mathrm{eff}$ value (as measured in \S \ref{fiduc_teff}) via interpolation from the above grid. The common-mode spectrum is subtracted from the flare-only spectrum of each integration. Next, a Planck function is fit to the common-mode corrected spectrum and subtracted to remove the remaining large-scale structure. Further discussion on this Planck subtraction step may be found in the Appendix. A summed output spectrum is generated by taking the average of the now-detrended spectra of all integrations, and uncertainty in the mean spectrum is computed using mean bootstrap resampling of the detrended spectra with replacement. The uncertainty in the mean of the 189 integrations of $\sim$30~ppm at 1.5$\mu$m and $R$=10 remains consistent with the expected precision of 35~ppm given by the JWST ETC. These uncertainties are smaller than the scatter in the mean spectrum itself, leaving uncorrected stellar noise as the primary limitation.

\section{Physical flare modeling and mitigation with RADYN}\label{rhd_modeling}
We perform a grid search comparing the flare spectrum of each integration with those predicted by 43 \texttt{RADYN} model sets that span four orders of magnitude in flare heating properties \citep{Kowalski:2024}. Each \texttt{RADYN} model evolves a 1D non-LTE stellar atmosphere in response to heating from a beam of non-thermal electrons by solving the coupled radiative-transport and hydrodynamics equations \citep{Allred:2006, Allred:2015}. Electron beams are described by three parameters: the peak electron flux in erg\,s$^{-1}$\,cm$^{-2}$, cutoff energy in keV, and index of accelerated electrons $\delta$ that determines the relative number of high and low-energy electrons (i.e. $n(E)\propto E^{-\delta}$, \citealt{Dulk:1985}). As a result, each model set is given a designation such as mF13-85-3, where the peak flux is 10$^{13}$\,erg\,s$^{-1}$\,cm$^{-2}$, the cutoff energy is 85 keV, and $\delta$=3. Further details are given in the Appendix.

Integrations during the six flare peaks of 3000--5000~K (top, \autoref{fig:activity_overview}) provide an ideal starting point for our \texttt{RADYN} exploration due to their high SNR and clear overlap with the \texttt{RADYN} grid's 4000--20,000~K flare $T_\mathrm{eff}$ range. Fits to these peaks inform the less straightforward analysis of the flare spectra during integrations of lower $T_\mathrm{eff}$ where quiescent features also become apparent (bottom, \autoref{fig:activity_overview}). The high SNR and overlap in flare $T_\mathrm{eff}$ of the flare peak integrations therefore give the clearest constraints on the physical properties of the electron beam. The peak line and continuum spectral features of each flare are shown in \autoref{fig:peak_spectra}, which enable independent tests of the beam properties from the lines and continuum of the same flare. A grid search of time-averaged spectral models from 1--10\,s are convolved to NIRISS resolution and fit to the H$\alpha$, P$\alpha$, P$\beta$, and the He$_\mathrm{I}$ IRT lines with the filling factor as a free parameter (bottom, \autoref{fig:peak_spectra}). Best-fit model sets, timestamps, and filling factors are measured for joint fits to all four lines. All model sets with residual errors within 100\% of the best-fit residual value are also recorded. No models with 10$^{10}$~erg~s$^{-1}$~cm$^{-2}$ beam fluxes or $>$85 keV electron beams produce residuals within 100\% of the best-fit residuals. 

A grid search of time-averaged and representative time-evolved spectral models from 1--10\,s are fit to the peak continuum of each flare. The grid search identifies the mF12-37-5 model set as the best fit to the peak flare spectra (top six panels, \autoref{fig:peak_spectra}), with filling factors of 0.46$\pm$0.2\% of the stellar surface and reduced $\chi^2$ values of 1.7--4.8. The mF12-17-5 set also provides a strong fit, with filling factors of 2.3$\pm$1.0\% and reduced $\chi^2$ values of 1.5--6.3. The grid search reveals a model deficiency at 0.82\,$\mu$m, where \texttt{RADYN} predicts a strong Paschen jump not supported by the data. Models with a smaller Paschen jump have higher electron fluxes ($\geq$7.5$\times$10$^{12}$ erg~s$^{-1}$~cm$^{-2}$), energies ($\geq$150 keV), or lower indices ($\leq$2.5) than the best-fit models and do not describe the observed spectra. We perform two-component fits to the lower-$T_\mathrm{eff}$ integrations, supplementing the \texttt{RADYN} continuum $f_{\lambda,\mathrm{RADYN}}$ with a quiescent contaminant spectrum $f_{\lambda,\mathrm{PHOENIX}}$ as described in the Appendix.

The \texttt{RADYN}-based transit spectroscopy mitigation pipeline begins with common-mode spectra computed by subtracting the best-fit model spectrum from each of the fiducial temperature spectra using \autoref{eq_1}, which are smoothed as in \S \ref{empirical_approach}. Common-mode spectra are interpolated to the flare $T_\mathrm{eff}$ of each integration and subtracted prior to fitting  \autoref{eq_1} to the detrended spectrum of the integration to produce a residual spectrum. When fitting \autoref{eq_1} solely for mitigation purposes at NIR wavelengths, we manually remove the Paschen jump discontinuity at 0.82\,$\mu$m and fit the spectra above and below 1.5\,$\mu$m separately as described in \citet{Howard:2023}. However, we do not modify the \texttt{RADYN} model outputs when fitting the flare peaks in order to predict X-ray or UV counterparts (\S \ref{escape_photochem}), as doing so invalidates prediction physicality.

\begin{figure*}
	\centering
	{
		\includegraphics[width=0.99\textwidth]{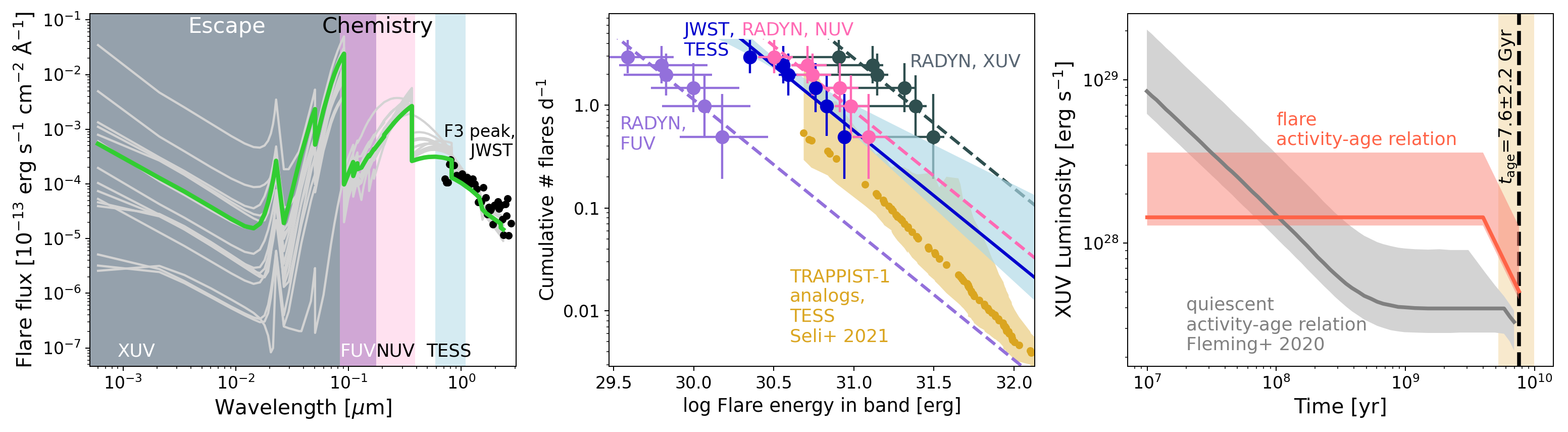}
	}
	\caption{Left: Best-fit \texttt{RADYN} model and uncertainty region for the F1 peak spectrum, determined from all models with residuals within 100\% of the best-fit value. The XUV emission driving atmospheric escape, FUV--NUV emission driving photochemistry, and TESS wavelengths are highlighted. Prediction capability degrades with distance from the NIR. Middle: \texttt{RADYN} spectral energy distributions enable scaling of the TESS-band flare rate into the XUV, FUV, and NUV. Right: Comparison of flaring and quiescent contributions to the XUV radiation environment, where the quiescent relation is reproduced from \citet{Fleming:2020}. The flaring XUV activity-age relation is constructed from the present-day XUV flare rate, estimated saturation flare rate relative to present-day \citep{Paudel:2018}, and estimated $\sim$4 Gyr saturation knee \citep{Fleming:2020}.}
	\label{fig:radyn_high_energy}
\end{figure*} 

\section{Stellar inputs for atmospheric escape and photochemistry with \texttt{RADYN} fits}\label{escape_photochem}
The high-energy counterparts to our peak flare spectra predicted by the best-fit mF12-37-5 model emit 8.9--28.9$\times$10$^{27}$\,erg\,s$^{-1}$ at XUV wavelengths (6--912~$\mathrm{\AA}$), 4.3--13.9$\times$10$^{26}$\,erg s$^{-1}$ at FUV wavelengths (912--1700~$\mathrm{\AA}$), and 3.4--11.4$\times$10$^{27}$\,erg\,s$^{-1}$ at NUV wavelengths (1700--3200~$\mathrm{\AA}$). Uncertainty regions are determined for the high-energy counterparts of each flare using an ensemble of time-averaged \texttt{RADYN} model predictions with fit qualities within 100\% error of the best-fit model value. The variation in predicted flux at each wavelength increases with distance from the NIR, with $\sim$60\%, 40\%, and 15\% variation observed for the XUV, FUV, and NUV, respectively. The best-fit \texttt{RADYN} models and uncertainty regions allow prediction of the scale factors between the TESS band and each high-energy counterpart. These scale factors are R$_\mathrm{XUV/TESS}$=3.6$_{-2.5}^{+0.6}$, R$_\mathrm{FUV/TESS}$=0.17$_{-0.08}^{+0.16}$, R$_\mathrm{NUV/TESS}$=1.4$_{-0.3}^{+0.4}$, which enable us to derive the contribution of flares to the stellar inputs driving atmospheric escape and photochemistry. We caution against over-interpreting X-ray predictions for specific flares as \citet{Tristan:2023} observe significant variation in the X-ray-to-optical relations of M0V events on a flare-by-flare basis. Our FUV-to-NUV ratio, $\mathcal{R}_\mathrm{FUV/NUV,~JWST}$=0.12$_{-0.06}^{+0.12}$, is slightly lower than that derived from simultaneous GALEX observations of earlier M dwarf flares, $\mathcal{R}_\mathrm{FUV/NUV,~GALEX}$= 0.46$\pm$0.23 \citep{Berger:2024}. This FUV-to-NUV ratio is also much smaller than the $\mathcal{R}_\mathrm{FUV/NUV,~JWST}>$1 predicted by the extreme electron beams required to reproduce short-wavelength superflare spectra in \citet{Kowalski:2025}.

We compute the cumulative flare frequency distribution (FFD) in the 0.6--1\,$\mu$m TESS band as described in the Appendix, in order to derive the X-ray and UV flare rates predicted by the best-fit \texttt{RADYN} models. The TESS-band FFD and TESS-to-XUV scaling of the best-fit \texttt{RADYN} model predicts a flare contribution of 1.6$\times$10$^{35}$\,erg to the cumulative XUV emission of TRAPPIST-1 per yr, 1.35$_{-0.15}^{+2.0}\times$ the quiescent emission of 3.75$\times$10$^{27}$\,erg\,s$^{-1}$ \citep{Wilson:2021} integrated over 1 yr. We assume the \citet{Fleming:2020} quiescent XUV--age relation for TRAPPIST-1 and a piecewise flaring XUV--age relation described below in order to estimate the relative flare contribution to the total XUV radiation history of the TRAPPIST-1 system. We find a cumulative 2.7$_{-0.3}^{+4}\times$ greater contribution from flaring, where the flaring XUV--age relation is constructed assuming saturated flaring of 4.5$\times$10$^{35}$\,erg\,yr$^{-1}$ below a spin-down threshold of 4 Gyr, and a linear decrease in flaring down to the current \texttt{RADYN} value. The relative decrease between the saturated and current flare rates of 2.8$\times$ is estimated from the difference in K2-measured flare rates between TRAPPIST-1 and very young ultracool dwarfs from \citet{Paudel:2018}.

\section{Injection and recovery tests for sensitivity to atmospheric signal strength}\label{inj_recov}
\begin{figure*}
	\centering
	{
		\includegraphics[width=0.99\textwidth]{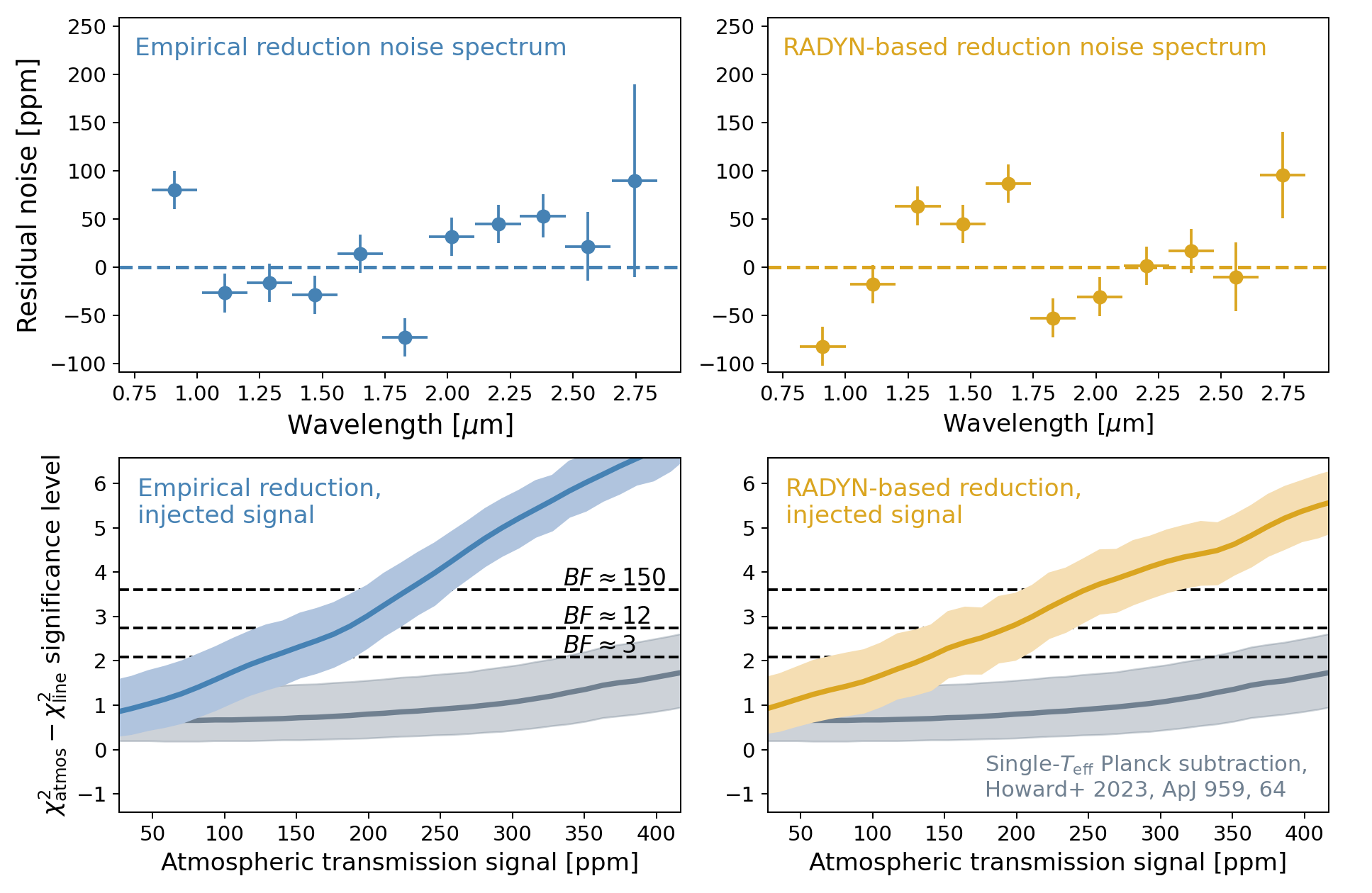}
	}
	\caption{Top: Residual flare contamination of 60--140 ppm in the $R$=10 average detrended spectrum of all 189 integrations following application of the empirical (left, blue) and \texttt{RADYN} model-based (gold, right) pipelines. Bottom: Detection significance versus transmission signal strength for the detrended spectra, derived from injection-and-recovery tests of \citet{Lustig-Yaeger:2019} CO$_2$ atmospheric transmission signals scaled from 12--480 ppm. Estimated Bayes Factors of 3, 12, and 150 represent ``weak," ``moderate," and ``strong" detections \citep{Trotta:2008}. The detection significance curve using the simple Planck fit approach from \citet{Howard:2023} is shown for reference (gray).}
	\label{fig:residspecs_injectiontests}
\end{figure*}

An average post-correction spectrum is produced from all 189 (5.5 hr) integrations for both the empirical and physics-based detrending procedures. The best-fit model to the flare-only spectrum is subtracted from each integration's spectrum and the mean spectrum of the residuals is computed for each procedure. The uncertainty in the mean residual spectrum is estimated from 200 bootstrap trials recomputing the mean with replacement. We then bin the residual detrended spectrum to $R$=10 in order to maximize signal while preserving noise structure at similar scales to those of expected transmission features (e.g., \citealt{Lustig-Yaeger:2019, Lim:2023, Radica:2024b}). The resulting detrended mean spectrum is shown for each reduction pipeline in the top panels of \autoref{fig:residspecs_injectiontests}. 

\begin{figure*}
	\centering
	{
		\includegraphics[width=0.99\textwidth]{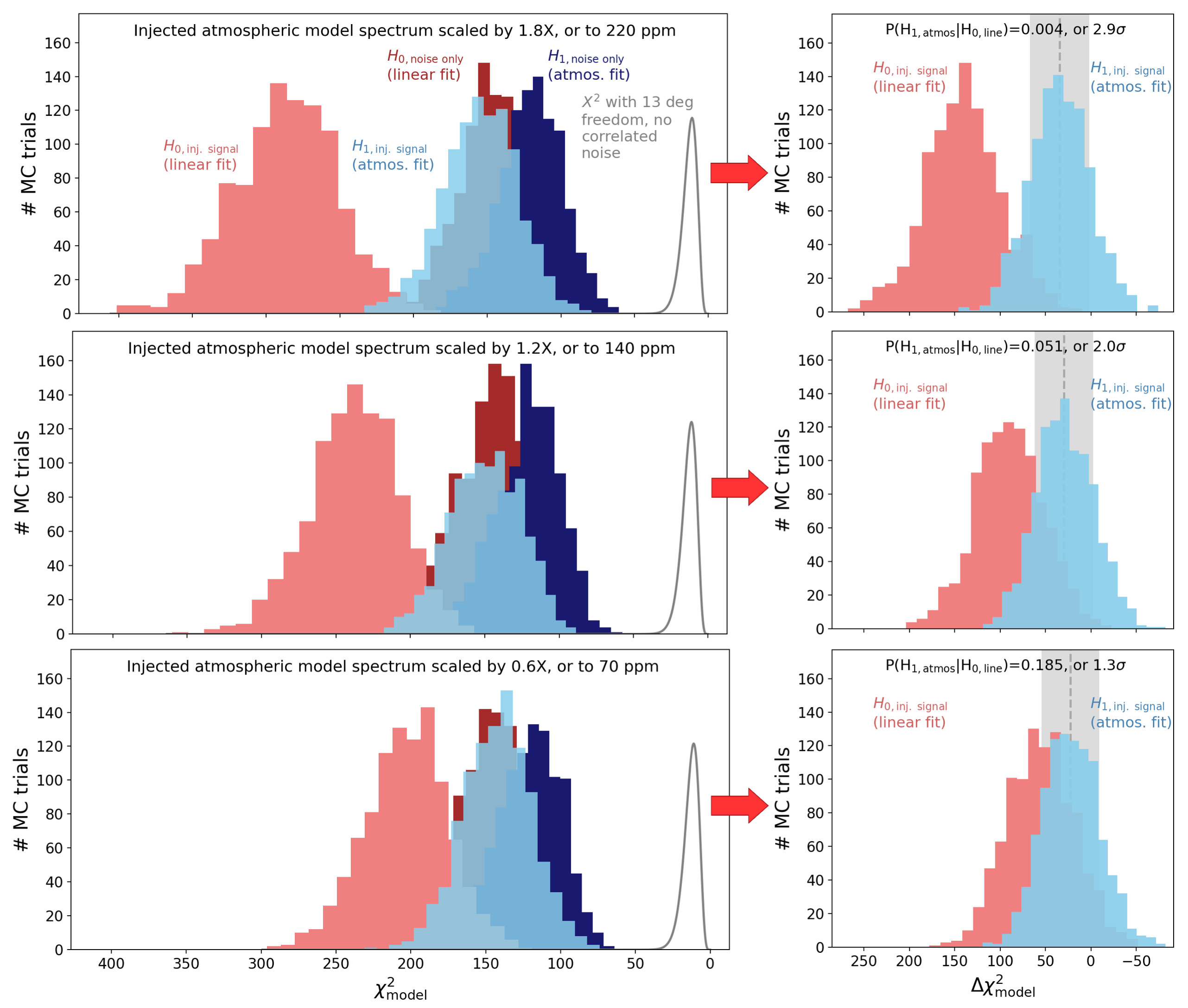}
	}
	\caption{Illustration of transmission spectrum injection and recovery testing procedure using measured $\chi^2$ (left panels) and $\Delta \chi^2$ (right panels) distributions. The \citet{Lustig-Yaeger:2019} TRAPPIST-1b 10 bar CO$_2$ atmospheric transmission spectrum is scaled by 0.6, 1.2, and 1.8$\times$ and injected into the pre-correction spectrum of all integrations. Bootstrapped $\chi^2$ values are recorded for linear and model atmosphere fits to the detrended mean flare spectrum under four scenarios: (i) linear model fit to the spectrum with the injected signal, (ii) atmosphere model fit to the spectrum with the injected signal, (iii) linear model fit to the spectrum in the absence of any injected signals, and (iv) atmospheric model fit to the spectrum in the absence of any injected signals. The presence of correlated noise in the no-atmosphere samples is highlighted by comparison to the random-variable $\chi^2$ function with the same 13 degrees of freedom. Hypothesis tests for atmosphere vs. linear fits are therefore computed using $\Delta \chi^2$ distributions where the correlated noise baselines have been subtracted.}
	\label{fig:deltachisq_hypothesis_tests}
\end{figure*}

Noise features are present in the detrended mean spectrum at levels up to those expected for secondary atmospheres, although the average noise beyond 1\,$\mu$m does not exceed 80~ppm. The largest feature appears at 1.8--2\,$\mu$m with amplitudes of 100 and 140~ppm for the empirical and physics-based reductions, respectively. We compute characteristic noise signals of 54$\pm$14 and 65$\pm$17 ppm beyond 1\,$\mu$m in the residual spectra using the mean successive difference (MSD; \citealt{Kamat:1953}) statistic. This is because variations in point-to-point scatter without regard to trend lines are most relevant to transmission feature confusion at very low resolution, and because mean or median absolute deviation (MAD) statistics do not sufficiently reflect the importance of larger but rarer noise features. We estimate the residual noise at the 4.3~$\mu$m CO$_2$ feature to be 160$\pm$140~ppm by extrapolating the best-fit \texttt{RADYN} models for each integration and correcting the \texttt{RADYN} extrapolations using the empirical pipeline prior to computing the average spectrum.

We use transmission spectrum injection and recovery tests to quantify detection sensitivity following the application of our empirical and physics-based detrending pipelines. For each test, a model 10~bar CO$_2$ transmission spectrum of TRAPPIST-1b \citep{Lustig-Yaeger:2019} is scaled in amplitude by test grid factors of 0.1--4 at 0.1 resolution and binned to $R$=10. These values correspond to scaling the $\sim$120 ppm characteristic signal strength from 12--480 ppm at 12 ppm resolution, where 120 ppm is the uncertainty-weighted average of the 2.0 and 2.8$\mu$m CO$_2$ feature amplitudes at $R$=10. The scaled transmission spectrum is injected into the flare spectrum of each integration prior to applying any detrending steps, and then the spectra are detrended with the empirical and physics-based pipelines. The mean residual spectrum is computed across all 189 integrations and the mean uncertainty is computed with 200 bootstrap resampling trials with replacement.

Next, $\chi^2$ values are computed for the best-fit linear and atmospheric model spectrum, where mean and 1$\sigma$ $\chi^2$ measurements are obtained from 1000 bootstrap trials varying the mean spectrum within its uncertainties, $\chi^2_\mathrm{inj.~signal,~line}$ and $\chi^2_\mathrm{inj.~signal,~atmos}$. The presence of correlated noise in the observed spectrum can bias the apparent $\chi^2$ values, so we also compute 1000 bootstrap trials to obtain mean and 1$\sigma$ $\chi^2$ measurements for the linear and atmospheric model fits in the absence of an injected atmosphere, $\chi^2_\mathrm{noise,~line}$ and $\chi^2_\mathrm{noise,~atmos}$. These enable us to obtain $\Delta \chi^2$ distributions where the effects of the correlated noise have been mitigated, $\Delta \chi^2_\mathrm{inj.~signal,~line}$=$\chi^2_\mathrm{inj.~signal,~line}$ - $\chi^2_\mathrm{noise,~line}$ and $\Delta \chi^2_\mathrm{inj.~signal,~atmos}$ = $\chi^2_\mathrm{inj.~signal,~atmos}$ - $\chi^2_\mathrm{noise,~atmos}$. The degree to which the injected atmospheric model is favored over a linear fit is determined from hypothesis testing using the $\Delta \chi^2$ distributions, where the $H_0$ distribution is given by $\Delta \chi^2_\mathrm{inj.~signal,~line}$ and $H_1$ by the mean and 1$\sigma$ values of $\Delta \chi^2_\mathrm{inj.~signal,~atmos}$ as shown in Fig. \ref{fig:deltachisq_hypothesis_tests}. We note $\chi^2$ testing of a preference for an atmospheric to a linear model is similar to the one-parameter atmospheric retrieval described in \S 3 of \citet{Crossfield_Kreidberg:2017} for H$_2$O feature amplitudes in young sub-Neptunes, and also similar to the analysis in \S 3.1 of \citet{Glidden:2025} for recent NIRSPEC transit observations of TRAPPIST-1. In particular, we follow \citet{Glidden:2025} in noting that a simple $\chi^2$ test for a linear fit is a helpful first step prior to performing the more comprehensive set of standard atmospheric retrievals, which we leave for future work.

Finally, we repeat the entire process 100$\times$ to verify the resulting atmospheric transmission signal versus significance curve converges at the 10\% level. The resulting significance curves for the empirical and physics-based detrending procedures are shown in \autoref{fig:residspecs_injectiontests}. A 3$\sigma$ detection is identified when $\leq$0.3\% of the $\chi^2~H_0$ distribution exceeds the expected $H_1$ values. The bootstrapped injection testing supports 3$\sigma$ detections for atmospheric transmission signals of 200$_{-50}^{+30}$ and 210$_{-60}^{+40}$~ppm with our empirical and physics-based reductions, respectively. Next, we estimate Bayes Factors (BF) from the $\chi^2$ significance results to assess the smallest transmission signals likely to produce moderate evidence for an atmosphere. We avoid $\sigma$-inflation effects by adopting the \citet{Sellke:2001} BF--$n_\sigma$ relation, which provides the most conservative BF estimate from among the relations explored in \citet{Kipping:2025} when inverted to estimate BF from $n_\sigma$, $n_\sigma$--BF. Following \citet{Trotta:2008}, we adopt BF=3, BF=12, and BF=150 for ``weak," ``moderate," and ``strong" detections, respectively. Given these assumptions, our empirical pipeline supports weak, moderate, and strong detections at 130$_{-60}^{+60}$, 190$_{-60}^{+40}$, and 230$_{-30}^{+30}$~ppm, respectively. Similarly, our RADYN-based pipeline supports weak, moderate, and strong detections at 140$_{-70}^{+60}$, 190$_{-60}^{+50}$, 250$_{-40}^{+60}$~ppm, respectively, within the errors of the empirical results. We emphasize these values are only applicable within the NIRISS wavelength range and may differ for $\geq$2.8~$\mu$m.

\section{Discussion and Conclusions}\label{discuss_conclude}
Although challenges remain, residuals of 54$\pm$14 and 65$\pm$17 ppm demonstrate that the large number of flare-contaminated transits do not necessarily have to be discarded. We emphasize the new $\sim$60 ppm residuals only hold for integrations during clearly-discernible flare peaks. Future work is required to extend these results to lower-level flaring, where signals are blended with other stellar contaminants. Furthermore, 70\% of our flare peak spectra occur out-of-transit, meaning higher residuals arising from TLS impacts (i.e. from flares as well as spots and faculae) are expected when attempting to mitigate flaring in-transit. One way our pipelines could be implemented into standard observer workflows would be to determine optimal flare spectra and uncertainties for in-flare integrations and subtract these spectra prior to combining the individual integrations and computing the transit spectrum. The results of \citep{Espinoza:2025} indicate the application of a GP framework for flare common-modes may also improve sensitivity to contaminated transit signals.

Improvements in model fidelity are also needed to understand contamination from spots, faculae, plages, and microflares on UCDs \citep{deWit:2024, Davoudi:2024, Berardo:2025}, as well as new methods for joint detrending of these stellar sources. Validation of sub-100 ppm residuals is also required for JWST transmission spectroscopy of other high-priority small planets around M-dwarf flare stars, e.g., LTT 1445, L 98-59, TOI 540, TOI 3884, and GJ 1132 \citep{Howard:2022a, Diamond-Lowe:2024, Murray:2025}, a focus of future studies. Our progress in flare modeling mirrors parallel efforts to improve agreement with observations for physics-based models of starspot and facular contamination with PHOENIX, MURaM, and MPS-ATLAS \citep{Waalkes:2024, Smitha:2025}, increasing the ultimate likelihood of obtaining robust constraints on secondary atmospheres from contaminated data.

The best-fit \texttt{RADYN} models for the observed line and continuum spectra of the flare peaks indicate moderate intensity electron beams (1--2$\times$10$^{12}$\,erg\,s$^{-1}$ cm$^{-2}$) and low cutoff energies ($\leq$40 keV) relative to the inferred beam properties during the peaks of many stellar flares \citep{Kowalski:2015}. Such relatively weak beams may help to explain TRAPPIST-1's lower-than-expected flare temperatures \citep{Maas:2022, Howard:2023}, drop in optical flare rate at bluer wavelengths \citep{Glazier:2020}, and lack of clear NUV flares observed with \textit{Swift} UVOT \citep{Becker:2020}. However, it remains premature to draw this conclusion in the absence of multiwavelength flare monitoring of TRAPPIST-1 relative to earlier M dwarfs, as two-temperature flare models \citep{Kowalski:2016} and changes in flare rate with time \citep{Wainer:2024} can also explain these trends. Improvements in \texttt{RADYN} model fidelity are needed to account for the presence of quiescent features and absence of a Paschen jump in the data.

Simultaneous observations at X-ray, UV, and NIR wavelengths are needed to confirm \texttt{RADYN} predictions of the high-energy counterparts to the TRAPPIST-1 flares. Such observations will improve the robustness of models for atmospheric escape and photochemistry in the TRAPPIST-1 system. Similarly, simultaneous millimeter observations probe (gyro)synchrotron emission in flares \citep{MacGregor:2021}, providing direct confirmation of the electron beam properties inferred by \texttt{RADYN}. Coordinated X-ray and millimeter observations during JWST observations of TRAPPIST-1b and e (GO 6456, 9256; PI: Allen) are being obtained with XMM Newton (PR 136870, PI: Howard), ALMA (2024.A.00040.S, PI: MacGregor), and potentially other facilities. Beyond TRAPPIST-1, the techniques outlined in this work form the basis for \texttt{RADYN} modeling of upcoming NIRISS observations of hundreds of M-dwarf flares from five young M0--M4 dwarfs (GO 7068; PI: Doshi). This program will provide stringent \texttt{RADYN} model fidelity tests at IR wavelengths and determine whether TRAPPIST-1 flares are typical of earlier M dwarf flares.

\section{Acknowledgements}\label{sec:acknowledge}
We would like to thank the anonymous referee for taking the time to review; your report genuinely helped improve the work. WH thanks Zach Berta-Thompson for a listening ear and suggesting the use of $\Delta \chi^2$ tests for \S \ref{inj_recov}. He also thanks Ryan MacDonald for pointing to the correct Bayesian evidence tables. WH also acknowledges the birth of Madelyn Howard, whom he hopes will one day share his sense of wonder at the Universe.

Funding for this work was provided by NASA through the NASA Hubble Fellowship grant HST-HF2-51531 awarded by the Space Telescope Science Institute, which is operated by the Association of Universities for Research in Astronomy, Inc., for NASA, under contract NAS5-26555. This work is based on observations made with the NASA/ESA/CSA James Webb Space Telescope. The data were obtained from the Mikulski Archive for Space Telescopes at the Space Telescope Science Institute, which is operated by the Association of Universities for Research in Astronomy, Inc., under NASA contract NAS 5-03127 for JWST. The specific observations analyzed can be accessed via \dataset{10.17909/sf2p-1k55}. These observations are associated with programs GTO 1201 and GO 2589. The authors acknowledge the Lim team for developing their observing program with a zero-exclusive-access period. This project was undertaken with the financial support of the Canadian Space Agency. Support for program number JWST-AR-05370 was provided through a grant from the STScI under NASA contract NAS5-03127. GTM acknowledges support from the National Science Foundation MPS-Ascend Postdoctoral Research Fellowship under Grant No.2402296

\software{\texttt{jwst} v1.12.5 \citep{bushouse_2023_10022973}, \texttt{exoTEDRF} \citep{Feinstein:2023, Radica:2022}, \texttt{RADYN} \citep{Allred:2015, Kowalski:2024}}

\appendix

\setcounter{figure}{0}
\renewcommand{\thefigure}{A\arabic{figure}}
\setcounter{table}{0}
\renewcommand{\thetable}{A\arabic{table}}

\section{Joint transit and flare light curve fitting procedure}\label{transit_corrections}
Joint transit and flare fitting to the white-light light curves is performed using the \texttt{batman} \citep{Kreidberg:2015} and \texttt{llamaradas-estelares} \citep{Tovar_Mendoza:2022} light curve templates. We perform 10$^5$ Monte Carlo (MC) trials using normal distributions for the inferior conjunction, transit depth, flare peak time, flare duration, and flare amplitude using mean estimates for the transit parameters from \citet{Gillon:2017} and visual estimates for the flare parameters, with standard deviations of 0.001 d, 0.005, 0.001 d 0.01, and 0.0001 respectively. The quadratic limb darkening coefficients use a uniform prior of 0--1 and are fit following \S 3.5 of \citet{Howard:2023}. The transit fits to the NIRSpec data and all NIRISS data fixed orbital parameters are reproduced from \S 3.5 of \citet{Howard:2023}. Uncertainties in fitted transit shapes of 0.3--0.4\% are determined from the 1$\sigma$ confidence intervals for MC fit quality.

\section{Further details on fiducial and common-mode spectra}\label{sec:extra_cmode_info}
The choice of a Planck function imposes a specific spectral shape and can bias $T_\mathrm{eff}$ depending on contamination, noise, or deviations from a true blackbody (e.g., molecular features). The fitted temperature then dictates the correction via common-mode subtraction, which in turn reshapes the spectrum and reinforces the original $T_\mathrm{eff}$ estimate. As a result, the $T_\mathrm{eff}$ parameter of the secondary Planck fit to the integration's common-mode-subtracted spectrum has already been influenced by the first fit and no longer holds physical significance. We avoid circularity concerns arising from multiple applications of blackbody fitting by using only the pre-correction $T_\mathrm{eff}$ values for common mode construction and spectral subtraction inputs. The purpose of the secondary Planck fit is limited to spectral whitening or removal of the blackbody-like large-scale structure left over after common-mode subtraction. The dependence of the Planck function fit on just two parameters prevents the unintentional injection of small scale features that could be confused with planetary signals, motivating its use in the whitening step. The fit parameters of these secondary fits are therefore given no physical interpretation and are not used further.
\begin{figure*}
	\centering
	{
		\includegraphics[width=0.99\textwidth]{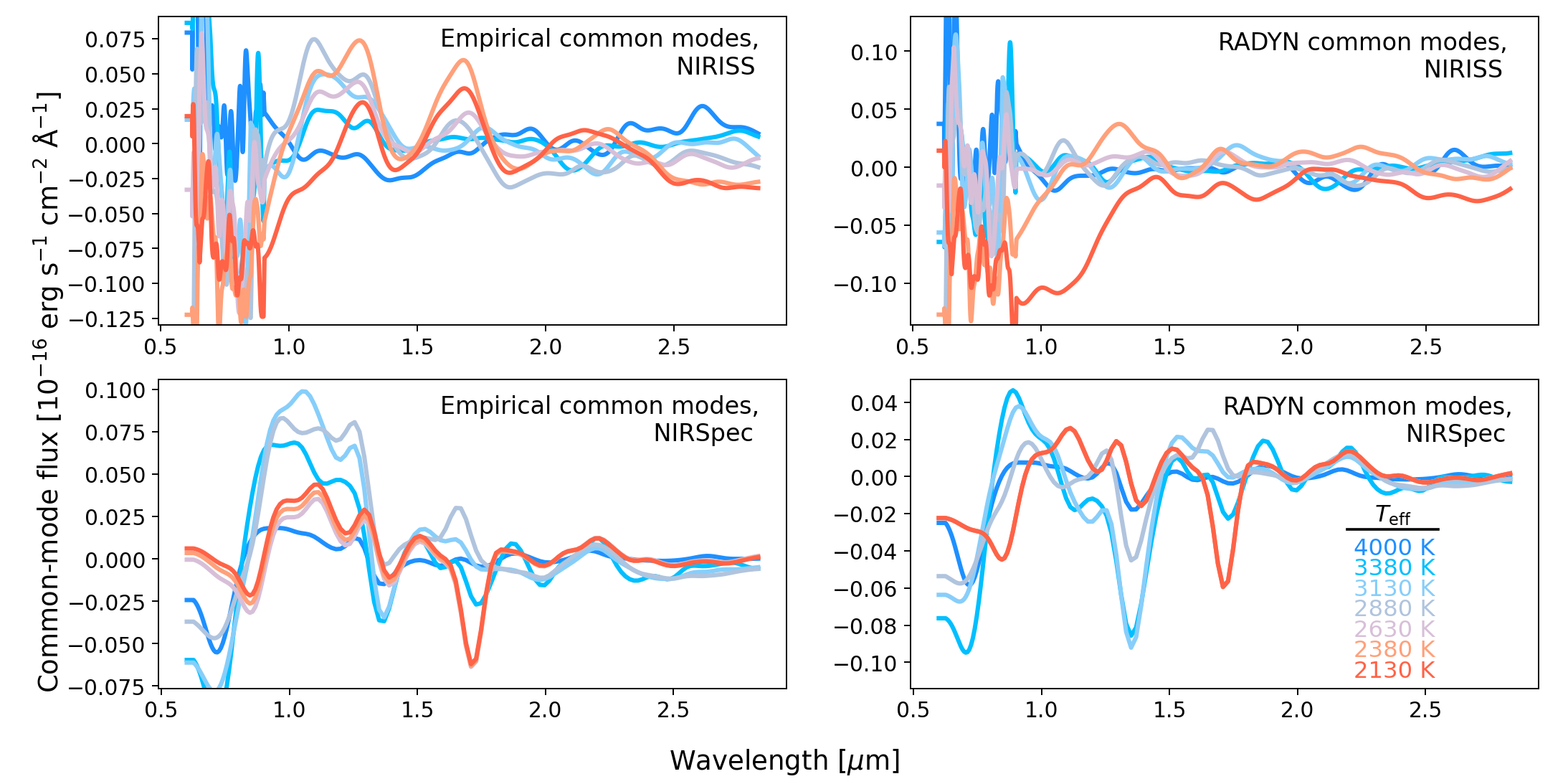}
	}
	\caption{Common-mode spectra computed from the fiducial temperature flare spectra, shown for each combination of NIRISS and NIRSpec and empirical and \texttt{RADYN}-based pipelines. The smooth evolution in common-mode spectra with temperature motivates our interpolation of common modes from the grid values. Also note the dip in the NIRSpec modes at $\sim$1.4$\mu$m}
	\label{fig:common_modes}
\end{figure*}

\begin{figure*}
	\centering
	{
		\includegraphics[width=0.92\textwidth]{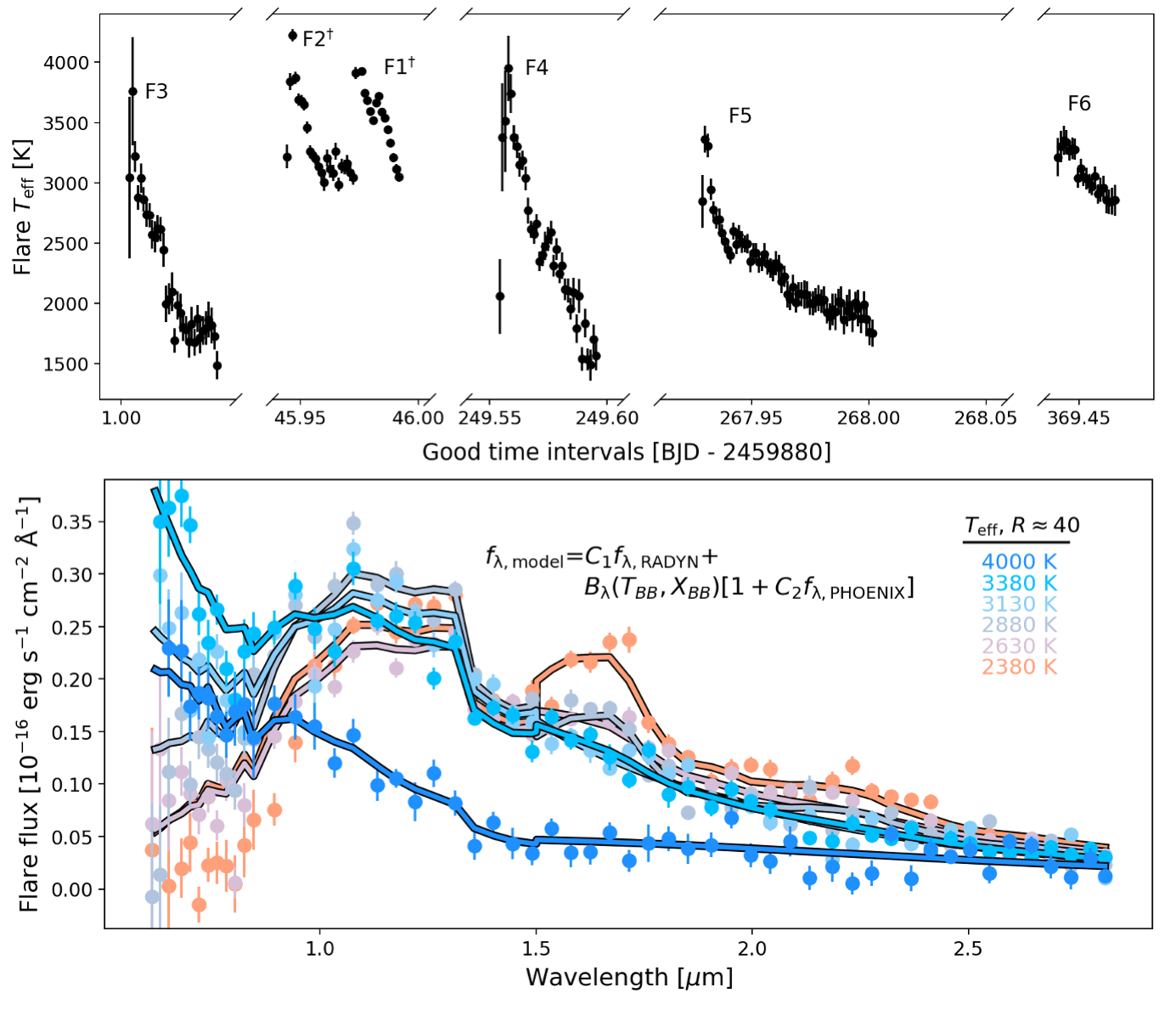}
	}
	\caption{Top: Flare $T_\mathrm{eff}$ time series for all 189 integrations in our GTI sample. Bottom: Combined RADYN and quiescent fits for a range of fiducial flare temperatures. We note the addition of the \texttt{PHOENIX} component is not intended to claim these features are physically correlated with the RADYN continuum, only that the combined fit better describes the flare spectrum when quiescent bleed-through is present.}
	\label{fig:addl_teff}
\end{figure*}

\section{Further description of the \texttt{RADYN} flare model grid}\label{sec:rhd_quiesc}

\subsection{\texttt{RADYN} grid setup and spectral outputs}\label{sec:rhd_setup}
The starting atmosphere is described by a uniform cross-sectional area, 10$^9$ cm loop half-length, 3$\times$10$^{10}$\,cm$^{-3}$ ambient electron density, $T_\mathrm{eff} \approx$ 3600\,K, log $g$ = 4.75, and 5\,MK gas temperature. The atmospheric response is calculated at a time resolution of 0.2\,s for an \citet{Aschwanden:2004} pulsed injection profile consisting of a 1\,s rise and 10\,s decrease in beam intensity. Spectral outputs for each model set are produced at 0.2\,s time resolution, including the flare continuum sampled at 95 points from 0.0006--4\,$\mu$m and line emission following the pressure broadening profiles of \citet{Tremblay_Bergeron:2009} and sampled at 0.0002\,$\mu$m resolution. \texttt{RADYN} lines overlapping the spectral range of JWST instruments include H$\alpha$, P$\alpha$, P$\beta$, and the He$_\mathrm{I}$ IRT. In addition to the \citet{Kowalski:2024} model grid, we follow the same procedure to compute a new model to fill out parameter space near initial tests of preferred models for the TRAPPIST-1 system, mF12-17-5. The new model was calculated for a 10$^{12}$\,erg\,s$^{-1}$\,cm$^{-2}$ beam flux, 17 keV cutoff energy, and $\delta$=5 using the \texttt{FP} code for Coulomb energy loss \citep{Allred:2020}.

\subsection{\texttt{RADYN} fits to low-temperature flare spectra with quiescent features}\label{sec:new_rhd_quiesc}
The quiescent contribution in low-$T_\mathrm{eff}$ integrations is incorporated into the RADYN fitting procedure via \autoref{eq_1}:
\begin{equation}
\label{eq_1}
    f_{\lambda,\mathrm{fl}} = C_1 f_{\lambda,\mathrm{RADYN}} + B_\lambda(T_\mathrm{BB},X_\mathrm{BB})[1+C_2 f_{\lambda,\mathrm{PHOENIX}}]. 
\end{equation}
Here, $C_1$ is a scale factor relative to the initial model value $X_\mathrm{fl}$=0.18\%, $C_2$ is a scale factor for the contaminant feature amplitude, $T_\mathrm{BB}$ is a free parameter for blackbody temperatures of 1000--5000~K, and $X_\mathrm{BB}\leq$0.5\% is a free parameter for filling factor. The spectral features of the \citet{Wilson:2021} PHOENIX model of TRAPPIST-1 are isolated from the overall shape of the SED by subtracting a heavily Gaussian-smoothed ($\sigma_G$=250) spectrum to produce $f_{\lambda,\mathrm{PHOENIX}}$. We find it necessary to multiply the $f_{\lambda,\mathrm{PHOENIX}}$ term by a single-$T_\mathrm{eff}$ blackbody function to account for the apparent modulation of quiescent feature amplitudes by the wavelength-dependent flare intensity. The \texttt{RADYN} continuum is constrained to lie within 1$\sigma$ of the mean for the 0.6--0.8\,$\mu$m observed spectrum to prevent unphysical solutions where the \texttt{RADYN} component is suppressed by the additional PHOENIX component, physically justified by the dominant nature of the flare continuum at shorter wavelengths. Unphysical solutions prefer electron energies of 350--500 keV in order to push the flare spectral peak below 0.6\,$\mu$m and NIR flux to zero. We therefore exclude the F10 and $>$85~keV model sets from consideration to enforce physical solutions since these are strongly disfavored by the line fits. The best-fit \texttt{RADYN} plus \texttt{PHOENIX} models are shown for a range of temperatures in the bottom panel of \autoref{fig:addl_teff}.

\section{An updated flare frequency distribution for TRAPPIST-1}\label{ffd_details}
The FFD describes the rate at which flares of energy $E_\mathrm{band}$ or greater are observed per day and is given by \autoref{eq_2}:
\begin{equation}
    \label{eq_2}
    \log{\nu_\mathrm{band}}=(1-\alpha)\log{E_\mathrm{band}} + \beta.
\end{equation}
Here, $\nu$ gives the number of flares with energy $\geq~E_\mathrm{band}$ per day, 1-$\alpha$ describes the frequency at which flares of various energies occur, and $\beta$ sets the overall flare rate. We report an FFD of $\log{\nu_\mathrm{TESS}}= -1.26_{-0.10}^{+0.53}\log{E_\mathrm{TESS}} + 38.96_{-2.92}^{+16.08}$ for our six flares and total monitoring time, where uncertainties on fit parameters are determined from 1000 MC posterior draws within the errors of the individual datapoints. Although the $\alpha=2.26_{-0.53}^{+0.10}$ power-law index appears to support smaller events as the dominant contribution of flares to the radiation environment ($\alpha>$2), the uncertainty range is also consistent with superflares as the dominant contributor ($\alpha<$2).

\bibliography{References.bib}
\end{document}